\documentstyle[preprint,aps,psfig]{revtex}

\def\mv{{\rm \bf m}}
\def\al{\alpha}
\def\alp{{\alpha^{'}}}
\def\avt#1{\left\langle{#1}\right\rangle}
\def\avg#1{\overline{#1}}

\begin{document}
\title{A MACROSCOPICALLY FRUSTRATED ISING MODEL}
\author{M. Pasquini$^{(1)}$ and M. Serva$^{(1,2)}$}
\address{$^{(1)}$Istituto Nazionale 
di Fisica della Materia, I-67010 Coppito, L'Aquila, Italy}
\address{$^{(2)}$Dipartimento di Matematica, Universit\`a 
dell'Aquila, I-67010 Coppito, L'Aquila, Italy}
\bigskip

\date{\today}

\maketitle

\begin{abstract}
A disordered spin glass model where both static and dynamical
properties depend on macroscopic magnetizations is presented.
These magnetizations interact via random
couplings and, therefore, the typical quenched realization of the system
exhibit a macroscopic frustration.
The model is solved by using a revisited replica approach,
and the broken symmetry solution turns out to coincide
with the symmetric solution. Some dynamical aspects of
the model are also discussed, showing how it could be
a useful tool for describing some properties
of real systems as, for example,
natural ecosystems or human social systems.
\end{abstract}
%]
%\narrowtext

\section{Introduction}
\bigskip

Macroscopic frustration can be found in different domains,
from interpersonal relationships to granular matter 
or natural ecosystems.
All these systems are characterized by 
frustrated components with a thermodynamically 
macroscopic size.
In other words, in all these systems, there are components 
whose size is comparable with that of the whole system
and which underly to the action of opposite forces. 
The classical example is the case of a man A who 
desires to spend some time with a dear friend B,
which, unfortunately, wants to bring his wife C,
which is really detested by A.

Dozen of examples can be found in nature. Consider
the antler of a deer, it is known that it represents
a frustrated phenotype. In fact,
sexual selection tends to prefer its growth in order to increment
the chance of reproduction but
antler is an obstacle in some situations, such a
predator pursuit in a forest and, therefore,
natural selection pressure is for its reduction.

From a more strict physical point of view, systems which exhibit 
frustration are very common (see \cite{KK} for
a general view).
For a disordered spin system Toulouse \cite{Toulouse}
has introduced the definition of frustration for an elementary plaquette 
of bonds, consisting in the product of the corresponding couplings.
Nevertheless, systems where frustration appears on macroscopic
scales are less ordinary and not yet investigated as far as we know.

In this paper we present a spin glass model where 
spins are organized in macroscopic sets, with the corresponding
macroscopic magnetizations interacting via random couplings.
For a typical random realization of the couplings,
the system is an ensemble of interacting 
frustrated macroscopic entities and, therefore,
it could be a natural candidate for mathematical modeling
of phenomena where macroscopic frustration plays a central role.

Let us briefly sum up the contents of the paper.

In sect. II the model is introduced.
The model becomes self-averaging when the number of components is large,
nevertheless, some considerations about its finite size version are
also written down.

In sect. III we look for a solution of the model 
using a revisited version of the replica trick.
This revised version could be applied in a more general context 
to a large class of models, as it will be explained.

In sect. IV and in sect. V, respectively,
the replica symmetry solution and the broken symmetry solution
{\it a la} Parisi are derived in detail. The two solutions
turn out to coincide, vanishing the benefits that 
the Parisi ansatz has in other spin glass models.

In sect. VI the symmetric solutions is studied in detail from 
a numerical point of view showing that,
at variance with the S-K model, it keeps its physical
meaning even at very low temperature.

In sect. VII some final remarks are contained, in particular
some dynamical aspects are illustrated.
Dynamics could be a profitable argument of future investigations
especially for its possible applications to ecosystems
and natural selection modeling.

\section{The model}
\bigskip

Let us consider an Hamiltonian where $N$ spins are divided in $L$ sets,
each set consisting of exactly $M=N/L$ spins. Each spin interacts 
with all other spins, but the coupling does not depend on 
the sites of the spins, but only on the sets of the spins involved.
In other words, two spins of different sets interact via a
coupling which depends only on the coordinates of the two sets
of membership.
Then, we can speak of coupling between sets rather then between 
spins. We also assume that spins of the same set do not interact.

This Hamiltonian can be written as

\begin{equation}
H_{M,L} (J,\sigma) = -\frac{1}{M\sqrt{L}}\sum_{k>l}
J_{k,l}\sigma_k\sigma_l  \ \ ,
\label{hamiltonian}
\end{equation}
where $J$ is a $N\times N$ symmetric matrix, consisting of $L^2$ blocks of
$M^2$ entries each, being $M$ the linear size of a block. All the $M^2$ entries
of a given block take the same value and, in particular, the diagonal blocks
consist of null entries. The free energy of the system is

\begin{equation}
f_{M,L} (J)= -\frac{1}{\beta M L} \ln \sum_{\{\sigma\}} 
\exp \left[ -\beta H_{M,L} (J,\sigma) \right]  \ \ ,
\label{fml}
\end{equation}
where the sum is intended over all the spin configurations.

The thermodynamic limit $N \to \infty$ can be obtained in two 
different ways since
$N$ is the product of two variables ($N=L M$). In fact,
the limit $L \to \infty$ would mean  to consider a system
whose properties and characteristics are the same
of those of the S-K model \cite{SK}. 
On the contrary, the limit $M \to \infty$ leads
 to
a mean field model with a macroscopic frustration.
The self-average properties are obtained by also performing
the limit $L \to \infty$  after the limit $M \to \infty$.
Nevertheless, non self-averaging macroscopic frustration also is
exhibited for finite $L$ as we will show later with an example. 
 
We thus perform the limit $M \to \infty$, keeping $L$ finite. 
After some algebra, the free energy reads
$$
f_L (J) = -\frac{1}{\beta L} \max_{\mv} \Gamma (J,\mv) \ \ ,
$$
where $\mv=(m_1,\dots,m_L)$, having defined the $i$th set magnetization $m_i$
as
$$
m_i = \lim_{M \to \infty} \frac{1}{M} \sum_{k \in i{\rm th \  set}} \sigma_k  
\ \ , 
$$
and where

\begin{equation}
\Gamma (J,\mv)=\frac{\beta}{\sqrt{L}} 
\sum_{i>j} J_{i,j} m_i m_j
+ \sum_i \Phi(m_i)  \ \ .
\label{gamma}
\end{equation}
The indices $i$ and $j$ run over the spin sets, and $J$ is now 
a symmetric $L \times L$ matrix, obtained from matrix in (\ref{hamiltonian})
substituting each block with a single entry,
being $J_{i,j}$ the value of the coupling  connecting a spin of set $i$
with a spin of set $j$, with $J_{i,i}=0 \ \forall i$.
Furthermore, $\Phi(m_i)$ represents the entropic term of spin set $i$
$$
\Phi(m_i)= -\frac{1+m_i}{2} \ln  \frac{1+m_i}{2}
-\frac{1-m_i}{2} \ln  \frac{1-m_i}{2} \ \ .
$$

Let us suppose that the non diagonal elements of $J$
are independent identically distributed random quenched variables.
For the sake of simplicity, we restrict ourselves to consider normal
Gaussian variables with vanishing average and unitary variance.
Our aim is to compute the quenched free energy
\begin{equation}
f= \lim_{L \to \infty}  f_L (J) 
= -\lim_{L \to \infty} \frac{1}{\beta L}
\ \avg{\max_\mv \Gamma(J,\mv)} \ \ ,
\label{f}
\end{equation}
where the last equality is due to the self-averaging
property of the free energy which holds in the large $L$ limit.
The max in (\ref{f}) is reached for $\mv^*=(m_1^*,\dots,m_L^*)$, 
which obey to the following $L$ self-consistent equations
\begin{equation}
m_i^*= \tanh \left[ \frac{\beta}{\sqrt{L}} 
\sum_{j} J_{i,j} m_j^* \right] \ \ \ \ 1\le i \le L \ \ .
\label{mi}
\end{equation}

We consider the large $L$ limit, because we have in mind 
a system with many macroscopic frustrated components,
nevertheless the glassy characteristics  (except
self-averaging) can be also found for finite $L$.
Consider, for instance $L=3$ with the product of the three couplings
with negative sign.
At low temperature (temperature below transition, not vanishing!) 
the system is degenerated since it has six different pure states
with the same free energy and with non trivial and
non all equal values of the three magnetizations involved. 

When  $L$ increases, frustration increases and also
the number of pure states corresponding to the same free energy.
We hope to find in this way an interesting spin glass model
with new peculiarities.

\section{Replica trick revisited}
\bigskip

In order to perform the limit $L \to \infty$ we need to compute
$\avg{\max_\mv \Gamma(J,\mv)}$. We will accomplish
this task by means of replica trick with a slight but crucial variant.
Let us stress from the beginning that this way of applying replica trick 
is not restricted to our model, but it is more general and,
in principle, could be of some help in solving many other 
models with macroscopic variables.
In fact, what we propose here is a useful technique for
computing  quantities of the type $\avg{\max_\mv \Gamma(J,\mv)}$,
i.e. an average whose argument is a maximum over an expression
which depends on random variables ($J$) and on variables 
to be maximized ($\mv$).

It is easy to check that

\begin{equation}
\avg {\max_m \Gamma(J,\mv)} =\lim_{\mu \to \infty} \lim_{n \to 0}
\frac{1}{\mu n}
\ln \avg{ \left[ \int d \mv \exp \left( \mu \Gamma (J,\mv) \right)
 \right]^n} \ \ , 
\label{rtrick}
\end{equation}
where $d \mv = \prod_i dm_i$.
In fact, after having performed the limit $n \to 0$ as in ordinary 
replica trick in right hand side of (\ref{rtrick}), the saddle point method
allows to compute the limit $\mu \to \infty$, giving equality (\ref{rtrick}).
The variable $\mu$ is here only an auxiliary one.
 
Making explicit the $n$ replicas, the average in right hand site of
(\ref{rtrick}) can be written as
$$
\avg { \left[ \int d\mv \exp \left( \mu \Gamma (J,\mv) \right)
\right]^n }
=\int \prod_\al d\mv^\al \exp G_n (\mu,\mv^1,\dots,\mv^n) \ \ ,
$$
having defined
$$
G_n (\mu,\mv^1,\dots,\mv^n)
\equiv  \ln \ \avg{ \ \exp  \sum_\al \mu \Gamma (J,\mv^\al) \ }
\ \ ,
$$
where the index $\al$ runs over the $n$ replicas.
Finally this leads to the following expression for the free energy $f$

\begin{equation}
f =- \lim_{L \to \infty} \lim_{\mu \to \infty} \lim_{n \to 0}
\frac{1}{\beta L \mu n} \ln \int \prod_\al d\mv^\al 
\exp G_n (\mu,\mv^1,\dots,\mv^n) \ \ .
\label{ftrick}
\end{equation}

In our case, taking in mind equation (\ref{gamma}), we can 
give an explicit expression for $G_n$.
For the sake of simplicity we do not write in the following  the argument
of $G_n$.
After have taking the averages over the Gaussian $J$ variables,
and after some algebra, one has
$$
G_n =
\frac{\mu^2 \beta^2}{4L} 
\sum_{\al, \alp}(\sum_i m_i^\al m_i^\alp)^2+
\mu \sum_{i,\al}\Phi(m_i^\al) \ \ ,
$$
where $\al$ and $\alp$ run over the replicas, 
and where terms not diverging with $L$ have been neglected 
since they would disappear in the successive limit
$L \to \infty$. By means of the parabolic maximum trick, 
the above expression can be rewritten as
$$
G_n =
\max_{\{q_{\al,\alp}\}} \left[ \frac{\mu^2 \beta^2}{2} 
\sum_{\al, \alp} \left( q_{\al,\alp}
\sum_i m_i^\al m_i^\alp -\frac{L}{2} q_{\al,\alp}^2 \right)
+\mu \sum_{i,\al}\Phi(m_i^\al) \right] \ \ ,
$$
where $\{q_{\al,\alp}\}$ is a $n\times n$ matrix, which represents
from a physical point of view the overlap between replicas in spin glass theory.

Now the integral in (\ref{ftrick}) can be fully factorized
among the different spin sets, individuated by the index $i$.
This fact allows us to perform the limit $L\to \infty$ which
gives the final expression for the free energy in the replica context:

\begin{equation}
f =-  \max_{ \{q_{\al,\alp}\}} \lim_{\mu \to \infty} \lim_{n \to 0}
\frac{1}{\beta \mu n} \ln \int \prod_\al dm^\al \exp \tilde{G}_n
\label{freplica}
\end{equation}
with
$$
\tilde{G}_n = 
 \frac{\mu^2 \beta^2}{2} 
\sum_{\al, \alp} \left( q_{\al,\alp}
 m^\al m^\alp -\frac{1}{2} q_{\al,\alp}^2 \right)
+\mu \sum_{\al}\Phi(m^\al) \ \ ,
$$
where now $m^1,\dots,m^n$ are $n$ replicas of a scalar magnetization.
Notice that interchange of the position between the $\max_{ \{q_{\al,\alp}\}}$ 
and the integration is allowed since in the limit $L \to \infty$ this maximum 
corresponds to a saddle point approximation of an integration 
with respect to the same variables $\{q_{\al,\alp}\}$.

\section{Replica symmetric solution}
\bigskip

In order to find a solution, i.e. to compute the
quenched free energy (\ref{freplica}), we start by trying the 
usual symmetry unbroken strategy. 
Let us stress that the diagonal terms of matrix $q$
are relevant for this model, at variance with the
celebrated replica solution of the S-K model \cite{SK}. 
Therefore, in spite of assuming that the diagonal terms vanish 
as in symmetry unbroken solution of S-K, we assume
$$
q_{\al,\alp}=
q+\frac{x}{\beta\mu}\delta_{\al,\alp} \ \ ,
$$ 
where $\delta_{\al,\alp}$ is the Kroeneker delta.
Notice that elements on the diagonal differ only 
for a quantity of the order of $\mu^{-1}$ from the other entries, 
otherwise one would have diverging terms in the limit $\mu \to \infty$. 
This fact implies that overlap turns out to be a constant
only once the limit $\mu \to \infty$ has been performed. 
With this choice one gets
$$
\tilde{G}_n = 
\left[ \frac{\mu^2 \beta^2}{2} 
q \left( \sum_{\al} m^\al \right)^2
+\frac{\mu\beta x}{2}\sum_{\al}\left( (m^\al)^2-q \right) \right]
+\mu \sum_{\al} \Phi(m^\al) \ \ ,
$$
where terms which vanish in the two limits 
$n \to 0$ and $\mu \to \infty$ have been neglected.
By means of the standard Gaussian trick we have

\begin{equation}
 \exp \left[ \frac{\mu^2 \beta^2}{2} 
q \left(\sum_{\al} m^\al \right)^2 \right] =
\avt {\exp \Big( \mu \beta \omega 
\sqrt{q} \sum_{\al} m^\al \Big) } \ \ ,
\label{gtrick}
\end{equation}
where the average $\avt{}$ is on an independent 
normal Gaussian variable $\omega$. The above  
equality allows for writing 
$$
\exp \tilde{G}_n =
\avt{ \prod_{\al}
\exp \Big[ \mu \beta \omega 
\sqrt{q} m^\al +\frac{\mu\beta x}{2} \left((m^\al)^2-q\right)
+\mu \Phi(m^\al) \Big] } \ \ .
$$

Notice that the argument inside the $\avt{}$ average
in the previous expression is fully factorized
among the $n$ replicas. For this reason the integral in (\ref{freplica})
becomes the $n$th power of a single integral, and therefore
the limit $n \to 0$ can be performed:
$$
f = -\max_{q,x} \lim_{\mu \to \infty}\frac{1}{\mu \beta} \
\avt{ \ \ln \int dm
\exp \left[ \mu \beta \omega \sqrt{q} m +\frac{\mu\beta x}{2}(m^2-q)
+\mu \Phi(m) \right] } \ \ .
$$
Finally, the limit $\mu \to \infty$ can be performed by means
of the saddle point technique, obtaining

\begin{equation}
f = -\max_{q,x} \ \avt{ \ \max_m \left[ \omega  \sqrt{q} m +
\frac{x}{2}(m^2-q) +\frac{\Phi(m)}{\beta} \right] } \ \ .
\label{fsym}
\end{equation}

Let us stress once again the important role played by the small symmetry breaking 
(non vanishing $x$) introduced in the overlap. In fact, if we fix $x=0$
choosing in this way a pure unbroken solution, the extremization 
with respect to $q$ would be impossible, since the argument in (\ref{fsym})
would diverge for $q \to \infty$.
It also should be noticed that at least one of the maximum with respect to
$q$ and $x$ could has become a minimum 
after having performed the limit $n \to 0$. 

\section{Failure of breaking}
\bigskip

Trying to apply the ordinary approach to spin glass models,
the following step consists in introducing an asymmetry 
in the overlap matrix. Assume now that 
$$
q_{\al,\alp} = q +\frac{x}{\beta\mu} \delta_{\al,\alp}
+\frac{y}{\beta\mu} \gamma_{\al,\alp} \ \ .
$$
Following Parisi parameterization \cite{P1,P2,P3,P4,MPV},
$\gamma_{\al,\alp}$ is a matrix whose entries vanish
except in $n/l$ quadratic blocks of $l^2$ elements along the diagonal,
where all entries are equal one. 
Notice that we have made explicit once again a factor $\mu^{-1}$,
otherwise we would have divergent terms.
In this case the maximum has to be taken with respect to
$q$,$x$,$y$ and $l$.

With this ansatz and neglecting terms vanishing
in the successive limits $n \to 0$ and $\mu \to \infty$,
$\tilde{G}_n$ turns out to be
$$
\tilde{G}_n = \frac{\mu^2 \beta^2}{2} 
q \left( \sum_{\al} m^\al \right)^2
+\frac{\mu\beta x}{2}\sum_{\al}\left( (m^\al)^2-q \right)
+\frac{\mu\beta y}{2}\sum_{k} \left[ \left( \sum_{\al \in k} m^\al
\right)^2-q l^2  \right]
+\mu \sum_{\al} \Phi(m^\al)  \ \ ,
$$
where the index $k$ runs over the $n/l$ blocks on the diagonal of $\gamma_{\al,\alp}$
and the sum on $\al \in k$ goes on the $l$ values of $\al$ corresponding to the 
$k$th block.

By means of the parabolic maximum trick it is possible to write
$$
\left[ \frac{\mu\beta y}{2} 
 \left(\sum_{\al \in k} m^\al \right)^2 \right] =
\max_{\rho_k} \left[ \sqrt{\mu \beta y} \ \rho_k \sum_{\al \in k} m^\al
- \frac{\rho_k^2}{2} \right] \ \ .
$$
In this way, repeating also the trick in (\ref{gtrick}),
we have factorized $\tilde{G}_n$ with respect to the $n/l$ blocks,
and, therefore, the limit $n \to 0$ can be performed.
One gets

\begin{equation}
f = -\max_{q,x,y,l} \ \lim_{\mu \to \infty}\frac{1}{\mu \beta l} 
\avt{ \ln \int \prod_\al dm^\al \ \max_\rho \hat{G}_n} \ \ ,
\label{f4}
\end{equation} 
with
$$
\hat{G}_n =\sum_{\al} \left[
\mu \beta \omega \sqrt{q} m^\al 
+\frac{\mu\beta x}{2} \left((m^\al)^2-q\right)
+\sqrt{\mu \beta y} \ \rho \ m^\al - \frac{\rho^2}{2 l}
-\frac{\mu\beta y}{2} q l +\mu \Phi(m^\al) \right] \ \ ,
$$
where now the index $\al$ runs over only a single block,
whose corresponds the scalar variable $\rho$, and
where $\avt{}$ means the average over the normal Gaussian $\omega$.

The $\max_\rho$ in (\ref{f4}) can be put outside the
integration. This change is allowed and 
can be understood by the same argument 
used after equation (\ref{freplica}). As a consequence, 
the integral in the previous expression 
is factorized among the $l$ replicas of a block,
and reduces to a single integral because of
the factor $l$ in the denominator. Moreover, this
integral can be computed by means of the saddle point
method in the limit $\mu \to \infty$, obtaining
$$
f = -\max_{q,x,y,l} \ \lim_{\mu \to \infty}\frac{1}{\mu \beta} 
\avt{ \max_{\rho,m} \left[
\mu \beta \omega \sqrt{q} m 
+\frac{\mu\beta x}{2} (m^2-q)
+\sqrt{\mu \beta y} \ \rho \ m - \frac{\rho^2}{2 l}
-\frac{\mu\beta y}{2} q l +\mu \Phi(m) \right] } \ \ .
$$
The maximum with respect to $\rho$ can be computed, and
then performing the limit $\mu \to \infty$ one
finally has
$$
f = -\max_{q,x,y,l} 
\avt{ \max_{m} \left[
\omega \sqrt{q} m +\frac{x + y l}{2} (m^2-q)
+\mu \Phi(m) \right] } \ \ .
$$

Unfortunately, this final result is exactly the same 
of the unbroken case (\ref{fsym}), the only difference being 
that the variable $x$ is substituted by $x+y l$, 
which is irrelevant when the maximum is taken.

This result could imply that the model simply has a constant overlap 
which depends only on the temperature; otherwise one should
admit that the Parisi ansatz for replica symmetry breaking is inappropriate
in this context.

\section{Understanding replica symmetric solution}
\bigskip

The unlucky result of the replica broken solution allows us to
suppose that the symmetric solution (\ref{fsym}) could be the exact solution
of the model. For this reason we have to study it in detail 
in order to get more evidences for supporting this hypothesis.

The extremization with respect to $q$, $x$ and $m$ 
(this last inside the average and, therefore,
for any different $\omega$) leads to a system of self-consistent equations:

\begin{equation}
\begin{array}{lll}
m_\omega=\tanh (\beta \sqrt{q} \omega +\beta x m_\omega) \\
q= \avt{\ m_\omega^2 \ } \\
x= \frac{1}{\sqrt{q}} \avt{\ \omega \ m_\omega \ }
\end{array}
\label{mstar}
\end{equation}
this system of equations is solved by $q^*$, $x^*$ and $m^*_\omega$ 
and the free energy may be written as
$$
f = - x^* q^* - \frac{1} {\beta} \avt{\ \Phi(m^*_\omega) \ }
$$
Let us stress that $q^*$ corresponds to a  maximum with respect to
$q$ while the limit $n \to 0$ has transformed $x^*$ in a minimum
with respect to $x$.

For a given $\omega$, the first equations (\ref{mstar}) which refer to 
the $m_\omega$ could have a single solution (a maximum) or 
three different solutions, depending on the temperature.
At low temperature we have a single solution for
$\omega \gtrsim x/\sqrt{q}$ and three solutions
for $\omega \lesssim x/\sqrt{q}$. 
Two of these correspond to a maximum and the third to a minimum,
and this introduce an element of uncertainty.
We follow the rule of taking the solution $m^*_\omega$
of the first equation which corresponds to the larger of the two maxima
for every given $\omega$. 

In fig. 1 we plot the overlap $q^*$ and the parameter $x^*$
as functions of the temperature $T\equiv 1/\beta$.
The spin glass transition occurs at the critical temperature
$T_c=2$, the same of the S-K model.

In fig. 2 the free energy $f$ and the entropy $S=\avt{\ \Phi(m^*_\omega) \ }$
are plotted as functions of the temperature $T$.
At $T=0$ the free energy is $f_0 = \sqrt{2/\pi} \simeq 0.798$,
which is very close to the value of the S-K symmetric solution.
On the contrary, the entropy simply vanishes at $T=0$ at variance with
the S-K case, where the negative entropy proves the unphysical
nature of the solution in that case.

Let us stress that how to take the right extreme point with respect to 
$q$, $x$ and $m$ is a crucial step of the solution, 
and our choice, previously described, could be inappropriate.
In fact, with the limit $n \to 0$ the maximum with respect to $x$
has become a minimum, and this could also has happened
for some of the $m_\omega$.
In this case one should look for the minimum
with respect to the $m_\omega$ (or for the second maximum), 
at least for a subset of $\omega$.
Indeed, at  this stage, we are not able to give a sure answer on this point,
which should be argument of future deep investigations.

\section{Conclusions}
\bigskip

A dynamical approach to our spin glass model could be of 
some help in deciding for the 
correct solution. Following equation (\ref{mi}),
the deterministic dynamics of $L$ magnetizations is
$$
m_i(t+1) = \tanh \left[ \frac{\beta}{\sqrt{L}} \sum_{j} J_{i,j}m_j(t) 
\right] \ \ \ \ 1\le i \le L \ \ .
$$
Let us remind that matrix $J$ has vanishing diagonal entries $J_{i,i}=0$,
so that at each step the new value of the individual magnetization $m_i$ 
does not depend on its previous one.
The above dynamics takes advantage of peculiar features. For instance,
at each updating it moves $m_i$ in a value corresponding to a minimum free energy 
with respect to $m_i$ itself keeping fixed the other magnetizations.
Moreover, the free energy always decreases
at each updating of a single magnetization.

The dynamics makes the system evolve toward a fixed point, which is a relative minimum 
of the free energy  (not, in general, a global minimum).
Repeating many times this evolution, starting
from different initial values for the magnetizations,
allows to find the global minimum corresponding to the solution of
the static spin glass model.  
Preliminary results seem to suggest that the 
theoretical symmetric solution of section VI
is slightly different from dynamic solution only for very low temperatures.
This not necessarily implies that 
the symmetric solution is not the correct one. In fact, in order to avoid
finite size effects, one has to deal with large lattices (large $L$) 
in numerical simulations, so that the basin of attraction of the global minimum 
tends reasonably to become so small that one never uses correct initial conditions
in spite of the large number of attempts.

The above mentioned features make 
such a dynamics for magnetization versatile and
very fast from a numerical point of view.
Furthermore, not only it is useful for understanding
the associated static model, but it is also interesting in itself.
In fact, it describes a dynamical system which monotonically relaxes towards
a stable point corresponding to a local minimum of the free energy.

For this reason it is the ideal candidate for modelizing
some complex systems, such as natural ecosystems, where
each agent or species try to maximize its own fitness in a given context
of other active agents. The fitness corresponds to the individual free energy
with changed sign (the part of the free energy which
depends on a given magnetization $m_i$), 
and the magnetization $m_i$ to the species degree of specialization.
The individual attempts to improve its own condition and it 
happens to push the whole systems to maximize the total fitness.
This is the very peculiar feature of many real systems
which is reproduced by our dynamical model,
which also exhibit other realistic peculiarities, such as
the fact that the phase space is a landscape of a large number
of local maxima of the fitness at low temperature. 
In case of  catastrophe (even a small change of the couplings)
the system is not anymore in a state of maximal fitness and
the evolution restarts 
towards a different local maximum (a new period
of stability in evolution story), which is not necessary
higher than the previous.

In conclusion, this model seems to be very versatile, since its dynamics
could become both a powerful benchmark where to test 
general hypothesis about spin glasses, and a paradigmatic model
for evolving complex systems.

\bigskip
\bigskip
\noindent
{\bf Acknowledgements}

The authors are indebted with Roberto Baviera
and Julien Raboanary.
MP  acknowledges the financial support {\it P.A.I.S. sez. G  
- L'Aquila} of I.N.F.M. 
MS is grateful to
Institute Superieure Polytecnique de Madagascar 
for hospitality.

\hfill\eject

\begin{figure}
\centerline{\psfig{file=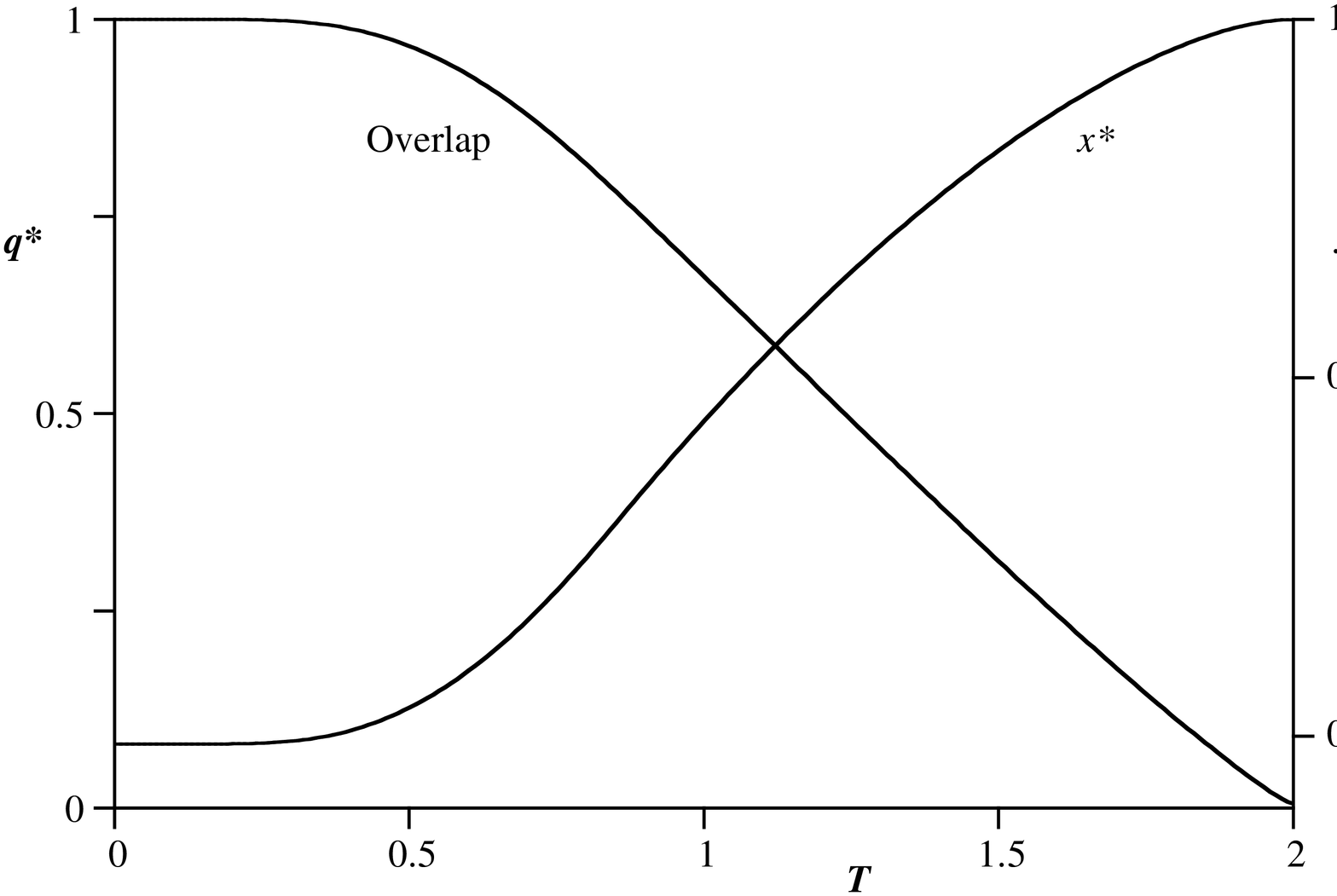,width=13.5cm,angle=270}}
\caption{
Overlap $q^*$ and parameter $x^*$ as functions of temperature $T$
for the symmetric solution.
The critical temperature below that we have a spin glass phase ($q^*>0$)
turns out to be $T_c = 2$.
}
\end{figure}

\begin{figure}
\centerline{\psfig{file=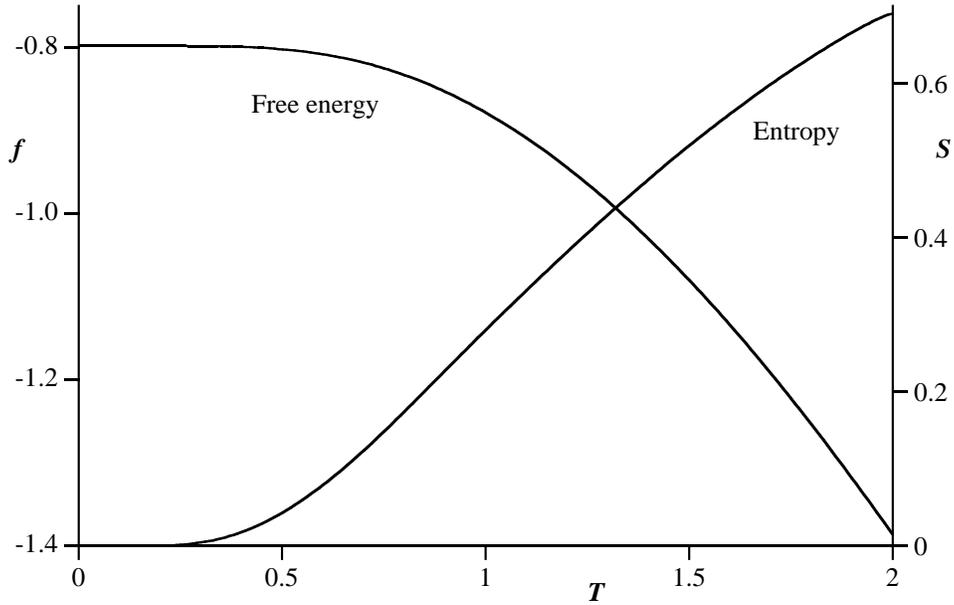,width=13.5cm,angle=270}}
\caption{
Free energy $f$ and entropy $S$ as functions of temperature $T$
for the symmetric solution.
In the limit $T\to 0$ the solution keeps a physical meaning
since the entropy never becomes negative.
}
\end{figure}


\begin{thebibliography}{99}

\bibitem{KK}
S. Kobe and T. Klotz,
Phys. Rev. E {\bf 52} (1995) 5660

\bibitem{Toulouse}
G. Toulouse,
Commun. Phys. {\bf 2} (1977) 99

\bibitem{SK}
D. Sherrington and S. Kirkpatrick,
Phys. Rev. Lett. {\bf 32} (1975) 1792

\bibitem{P1}
G. Parisi, Phys. Lett. A {\bf 73} (1979) 203 

\bibitem{P2}
G. Parisi, J. Phys. A {\bf 13} (1980) L115 

\bibitem{P3}
G. Parisi, J. Phys. A {\bf 13} (1980) 1101

\bibitem{P4}
G. Parisi, Phys. Rev. Lett. {\bf 50} (1983) 1946

\bibitem{MPV}
M. Mezard, G. Parisi and M. Virasoro, 
{\it Spin glass theory and beyond}, World Scientific Singapore (1988)

\end{thebibliography}
\end{document}